\documentclass[useAMS]{mn2e}

\usepackage{url,graphicx}
\usepackage{amsmath,amsfonts}
\usepackage{amssymb}

\def\reff@jnl#1{{\rm#1\/}}

\def\aj{\reff@jnl{AJ}}                  
\def\araa{\reff@jnl{ARA\&A}}            
\def\apj{\reff@jnl{ApJ}}                        
\def\apjl{\reff@jnl{ApJ}}               
\def\apjs{\reff@jnl{ApJS}}              
\def\ao{\reff@jnl{Appl.Optics}}         
\def\apss{\reff@jnl{Ap\&SS}}            
\def\aap{\reff@jnl{A\&A}}               
\def\aapr{\reff@jnl{A\&A~Rev.}}         
\def\aaps{\reff@jnl{A\&AS}}             
\def\azh{\reff@jnl{AZh}}                        
\def\baas{\reff@jnl{BAAS}}              
\def\jrasc{\reff@jnl{JRASC}}            
\def\memras{\reff@jnl{MmRAS}}           
\def\mnras{\reff@jnl{MNRAS}}            
\def\pra{\reff@jnl{Phys.Rev.A}}         
\def\prb{\reff@jnl{Phys.Rev.B}}         
\def\prc{\reff@jnl{Phys.Rev.C}}         
\def\prd{\reff@jnl{Phys.Rev.D}}         
\def\prl{\reff@jnl{Phys.Rev.Lett}}      
\def\pasp{\reff@jnl{PASP}}              
\def\pasj{\reff@jnl{PASJ}}              
\def\qjras{\reff@jnl{QJRAS}}            
\def\skytel{\reff@jnl{S\&T}}            
\def\solphys{\reff@jnl{Solar~Phys.}}    
\def\sovast{\reff@jnl{Soviet~Ast.}}     
\def\ssr{\reff@jnl{Space~Sci.Rev.}}     
\def\zap{\reff@jnl{ZAp}}                        
\def\nat{\reff@jnl{Nature}}             

\title[Candidate ultra-bright submillimetre galaxies]{A near/mid infrared search for  ultra-bright submillimetre galaxies: Searching for Cosmic Eyelash Analogues}

\author[S. Iglesias-Groth and  Other]{S. Iglesias-Groth$^{1,2}$\thanks{E-mail:
sigroth@iac.es},
A. D\'{\i}az--S\'anchez$^{3}$,
R. Rebolo$^{1,2}$
and
H. Dannerbauer$^{1,2}$\footnotemark[1]\thanks{This paper makes use of data from catalogues
VISTA and WISE} \\
$^{1}$ Instituto de Astrof\'\i sica de Canarias, La Laguna 38200, Spain\\
$^{2}$Departamento de Astrof\'\i sica de la Universiad de La Laguna, Avda. Francisco S\'anchez, La Laguna, 38200, Spain \\
$^{3}$Departamento F\'\i sica Aplicada, Universidad Polit\'ecnica de Cartagena, Campus Muralla del Mar, 30202 Cartagena, Murcia, Spain}

\begin{document}

\date{2016}

\pagerange{\pageref{firstpage}--\pageref{lastpage}} \pubyear{2016}

\maketitle

\label{firstpage}

\begin{abstract}
We present results from a near/mid IR search for submillimetre galaxies over a region of 
6230 sq deg. of the southern sky. We used a cross-correlation of the VISTA Hemispheric Survey (VHS) and the WISE database to identify bright galaxies (K$_s\leq$ 18.2)  with near/mid IR colours similar to those of the high redshift lensed sub-mm galaxy SMM J2135-0102. We find 7 galaxies which fulfill all five adopted 
near/mid IR colour (NMIRQC) criteria and resemble the SED of the reference galaxy at these wavelengths. For these galaxies, which are broadly distributed in 
the sky, we determined photometric redshifts in the range z=1.6-3.2. 
We searched the VHS for clusters of galaxies, which may be acting as gravitational lenses, and found that 6 out of the 7 galaxies are located within 
3.5 arcmin of a cluster/group of galaxies. Using the J-K$_s$ vs J sequences we determine photometric redshifts for these clusters/groups in the range z=0.2-0.9. 
We propose the newly identified sources are ultra-bright high redshift lensed SMG candidates. Follow-up observations in the sub-mm and mm are key to determine the ultimate nature of these objects. 
\end{abstract}


\section{Introduction}

Since their discovery (Smail et al. 1997; Hughes et al. 1998) the so-called submillimetre galaxies (SMGs) are a vital galaxy
population in order to understand the formation and evolution of massive galaxies in the distant universe (see for a review Casey et
al. 2014). These systems have extreme star formation rates of several hundred to thousand solar masses per year (e.g., Magnelli et
al. 2012), are molecular gas-rich with M$_{mol-gas}$= $\sim$ few times 10$^{10}$  M$_{\odot}$ (e.g., Greve et al. 2005; Bothwell et al. 2013) and the redshift distribution peaks $\approx$ $z=2.2-3.0$ (Chapman et al. 2005; 
Simpson et al. 2014; Miettinen et al. 2015; Strandet et al. 2016) depending on the selected wavelength in the (sub)mm window. Typical sizes of these dusty 
starbursts range between R$_{e}=0.6-2.0$ kpc (Ikarashi et al. 2015; Simpson et al. 2015; Hodge et al. 2016). Mergers and cold gas infall are potential agents 
of such star formation,  however the available studies have not yet established the main cause of the high star-formation rates (e.g., Swinbank et al. 2008; 
Hayward et al. 2013; Michalowski et al. 2012; Narayanan et al. 2015). 

Despite  intrinsic luminosities of L$_{bol}$$\geq$ few times 10$^{12}$ to 10$^{13}$  
L$_{\odot}$ (Magnelli et al. 2012; Ivison et al. 2013), these dusty galaxies are very challenging objects for observational studies in the near IR
(e.g., Dannerbauer et al. 2002, 2008; Younger et al. 2007) and mid infrared (e.g. Pope et al. 2008). The brightest unlensed SMGs have observed flux densities 
of up to $\sim$ 10 mJy at 850 $\mu$m (e.g., Karim et al. 2013). Gravitational lensing via massive galaxy clusters can enhance the apparent brightness of SMGs  
without altering their colours (Smail et al. 1997) and cluster surveys (e.g. Smail et al. 2002; Johansson et al. 2011) indeed led to the detection 
of SMGs with high amplification factors (30-40) like that of the outstanding galaxy SMM J2135$-$0102 at $z=2.3259$ (the Cosmic Eyelash, hereafter SMM J2135, 
Swinbank et al. 2010; Ivison et al. 2010; Danielson et al. 2011). The South Pole Telescope and the Herschel and  Planck space missions have unveiled similarly 
bright lensed SMGs (Negrello et al. 2010; Vieira et al. 2010, 2013;  Weiss et al. 2013; Ca\~{n}ameras et al. 2015; Harrington et al. 2016; Strandet et al. 2016). 
In spite of the amplification factors provided by lensing, these SMGs are still rather faint in the optical and near-IR as a consequence  of internal dust obscuration, 
and  their properties in this spectral range are not well understood yet (e.g. Dannerbauer et al. 2002, 2004, 2008; Dunlop et al. 2004; Younger et al. 2007; 
Walter et al. 2012; Simpson et al. 2014; Hodge et al. 2015; Dye et al. 2015). The spectral energy distribution  of  bright high redshift sources like SMM J2135$-$0102 (hereafter SMM J2135) 
shows a very steep increase in flux as we move from the optical to the near-IR and mid-IR which could potentially be  used to identify  other similar galaxies.

We took advantage of  the VISTA Hemisphere Survey (VHS)\footnote{\url{http://horus.roe.ac.uk/vsa/index.html}} which has already covered an area of more than 8000 sq. deg. of the southern sky at near-IR bands and the Wide-field Infrared Survey Explorer (WISE) mission archive database in the mid-IR in order to find brighter
analogues of SMM J2135.  We adopted as a reference the  near and mid-IR colours of this galaxy. 
Given the limiting magnitudes of the VHS/WISE surveys we expect the selected sample of similar colour galaxies will consist mainly of gravitationally lensed sources (Blain 1996, Negrello et al. 2007). Lensing of SMGs is strongly favored due to the steep number counts of this source population (Blain et al. 1996; Perrotta et al. 2002, 2003; Negrello et al. 2007; Negrello et al. 2010). Thus, if we conduct searches for sources with similar SEDs as known dusty starbursts in the near/mid IR wavelength 
regime, the above mentioned lensing bias should guarantee that the majority of the selected sources are massive galaxies with strong FIR/submm emission triggered by on-going intense star formation. Thus, the contamination by massive, early type galaxies should be rather negligible.

Our goal is to find the brightest high redshift (z$\sim$2) analogues of SMM J2135 in the sky such that follow-up studies can be 
performed in the optical, near/mid IR with moderate investment of observing time at the largest telescopes. Finding new examples of 
this important class of strongly lensed  star-forming galaxies should help  to determine their extinction properties in the 
visible/IR, perform extensive characterisation in the millimetre/submillimetre range and obtain a better understanding of how and where  stars are formed in these galaxies. Brighter analogues of the SMG  SMM J2135 in the redshift range z=2-3 will also allow us  to investigate star formation on 100 pc scales as a function of redshift using ALMA (Dye et al. 2015, Swinbank et al. 2015).

In the following sections we describe our search for bright SMGs in a large fraction of (approx. 30 \%) of the southern sky and the resulting candidate galaxies. In Section 2 
we present the infrared data from VISTA/VHS and WISE (Vega mag-system) used in the search, the photometry of the reference galaxy SMM J2135 and the adopted colour criteria  to identify potential analogues of this galaxy. Section 3 reports the results of the search and the identification of clusters of galaxies close to the  new candidate SMGs.

\begin{table*}
\begin{minipage}{170mm}
\begin{center}
\caption{ Optical and near/mid-IR fluxes of the sub-mm galaxy SMM J2135-0102 (z=2.3259) }
\label{tab:table_1}
\scriptsize{
\begin{tabular}{ccc}
\hline

Filter & Flux & Mag (Vega) \\

\hline
$V$ & 0.9 $\pm 0.2$  $\mu$Jy & 24.1  $\pm 0.2$\\
$I$ & 1.4 $\pm 0.4$  $\mu$Jy & 23.2  $\pm 0.2$\\
$J^1$ & & $\gtrsim $ 20.5 \\
$K_s$ & 36 $\pm 4$ $\mu$Jy & 18.00  $\pm 0.07 $\\
$[3.6]$ & 0.13 $\pm 0.02$ mJy & 15.82  $\pm 0.2 $\\
$[4.5]$ & 0.21  $\pm 0.02$ mJy  & 14.83  $\pm 0.1 $\\
$[8]$ &0.32  $\pm 0.05$ mJy & 13.2  $\pm 0.2 $\\
$[24]$ & 2.6  $\pm 0.2$ mJy & 8.6  $\pm 0.1 $\\
$W1^2$ & 0.12 $\pm 0.02$ mJy & 16.0  $\pm 0.2 $ \\
$W2^3$ & 0.23  $\pm 0.02$ mJy & 14.7  $\pm 0.1 $ \\
$W3^4$ & 0.24  $\pm 0.05$ mJy & 12.8  $\pm 0.2$\\
$W4$ & 2.5  $\pm 0.8$ mJy & 8.8  $\pm 0.3 $\\

\hline
\end{tabular}
}

$^1$ From the  VHS J-band image.

$^2$ Transformed from the Spitzer [3.6] band using the template SED of SMM J2135. 

$^3$ Transformed from the Spitzer [4.5] band using the template SED of SMM J2135. 

$^4$ Transformed from the Spitzer [8] band using the template SED of SMM J2135. 

\end{center}
\end{minipage}
\end{table*}

\begin{figure*}

\includegraphics[angle=0,width=18cm,height=4cm]{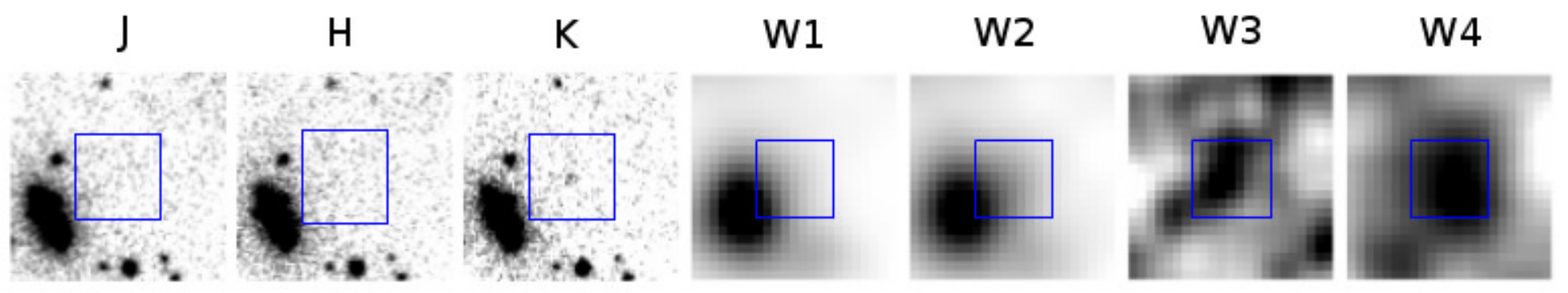}
\caption{Images 0.5$\times$ 0.5 arcmin$^2$ for SMM J2135-0102 in J, H, K$_s$ (VISTA/VHS) and  W1, W2, W3 and W4 (WISE) bands.
The marked boxes are 12 $\times$ 12 arcsec$^2$.
}
\label{fig:f1}
\end{figure*}

\begin{figure*}
\includegraphics[width=8cm]{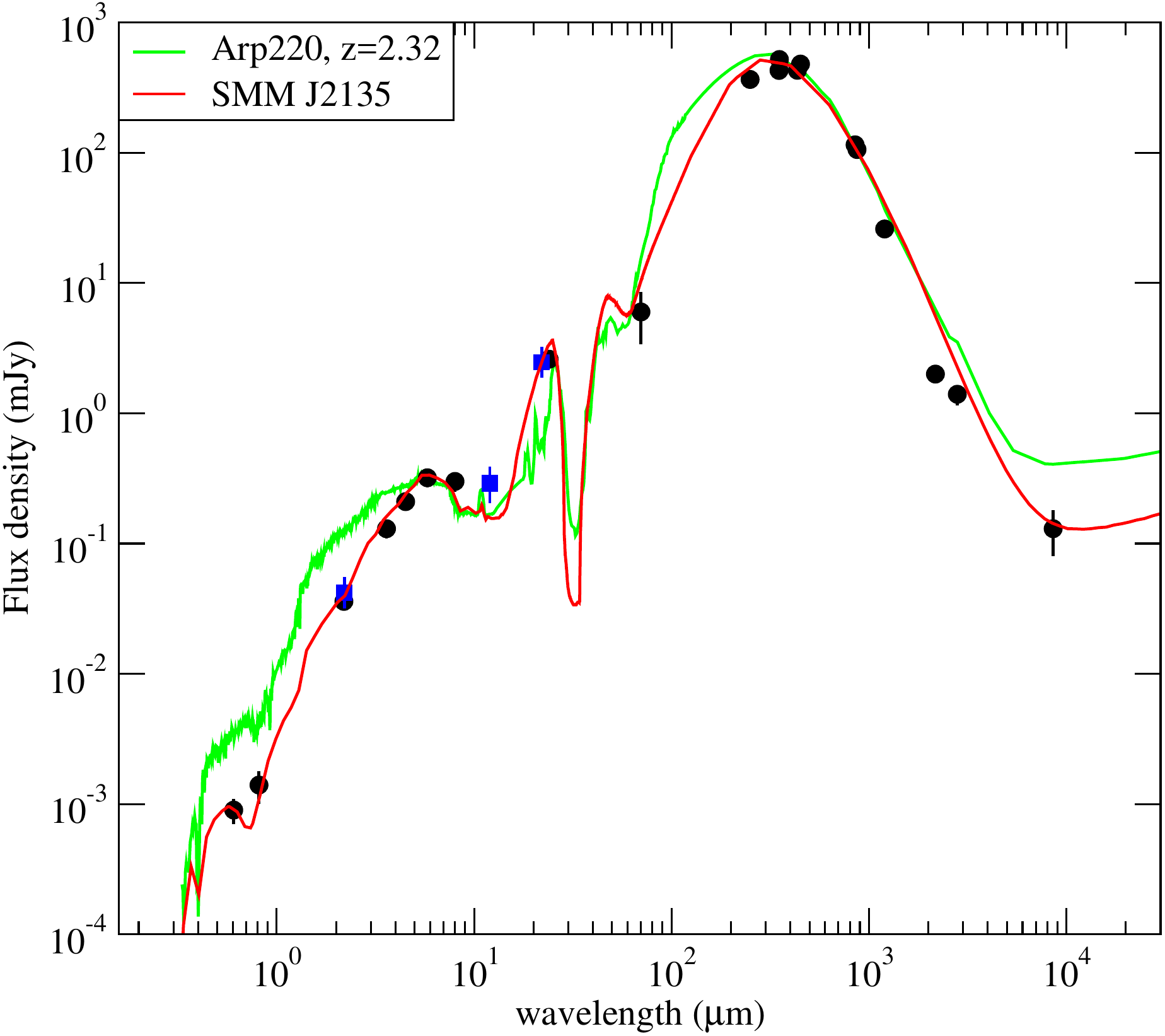}

\caption{
The Spectral Energy Distribution of  SMM J2135$-$0102.  The green line is the SED  of the well known
luminous IR galaxy Arp 220 redshifted to z=2.3 with appropriate normalisation and the red line is the GRASIL SED given in Lapi et al. (2011). Black points from Ivison et al. (2010) and blue boxes are our measurements.
}
\label{fig:f2}
\end{figure*}

\section[]{The Search}

 We will take  advantage of the large sky coverage  provided by the near-IR VISTA Hemispheric Survey (VHS) and the mid-IR
WISE mission to conduct a search for bright analogues of the SMG SMM J2135.

\subsection[]{The VISTA Hemispheric Survey: VHS}
The VHS is a near-infrared ESO public survey designed to map the entire Southern hemisphere
 in the J and K$_s$ broad band filters with average 5$\sigma$ depths of J = 19.5 $\pm$ 0.3 mag
  and K$_s$ = 18.5 $\pm$ 0.3 mag. In some particular areas also Y and H band observations are performed. The 4-m VISTA telescope (Emerson 2001; Emerson et al. 2004) operates since 2009 at ESO's Cerro Paranal Observatory 
  in Chile and has so far covered a sky area of about 8000 deg$^2$. It is equipped with a wide-field infrared camera VIRCAM (Dalton et al. 2006) composed of 16 Raytheon detectors 2048-2048 pixel array each, with a mean plate scale of 0.34 arcsec, giving a field of view of 1.65 degrees in diameter.

The VHS images are processed and calibrated automatically by a dedicated science pipeline implemented by the Cambridge
Astronomical Survey Unit (CASU). Standard reduction and processing steps include dark and sky subtraction, flat field
correction, linearity correction, destripe and jitter stacking. For a detailed description we refer to the CASU webpage
http://casu.ast.cam.ac.uk/surveys-projects/vista as well as to Irwin et al. (2004) and Lewis et al. (2010).
The photometry provided in the VHS catalogue is calibrated using the magnitudes of colour-selected 2MASS stars converted onto the VISTA system using colour equations including terms accounting for interstellar reddening. We used the catalogue J and K$_s$-band 1 arcsec aperture corrected (aperMag3) magnitudes  for the selection of  targets.

\subsection[]{The Wide-field Infrared Survey Explorer}
The NASA mission WISE completed an all-sky survey in four mid-IR bands (Wright et al. 2010) centred  at  3.4, 4.6, 12, and 22 $\mu$m (referred as W1, W2, W3 and W4, respectively) with an angular resolution of 6.1$Ó$, 6.4$Ó$, 6.5$Ó$, and 12.0$Ó$ and
5$\sigma$ point  source sensitivities of 0.07, 0.1, 0.86 and 5.4 mJy,  respectively.
The WISE photometric pipeline provides a number of quality flags useful in selecting point sources and extended
sources. We use the AllWISE Source Catalog PSF profile-fit measurements that are made by chi-squared minimization on the "stack" of all
Single-exposure frames in all bands covering a deep source detection
\footnote{\url{http://wise2.ipac.caltech.edu/docs/release/allwise/}}. Only measurements with S/N$>$3.0 will be considered.

\begin{table*}
\begin{minipage}{170mm}
\begin{center}
\caption{Infrared colour criteria of potential SMM J2135 analogue galaxies}
\label{tab:table_3}
\scriptsize{
\begin{tabular}{ccc}
\hline
Mag & Color & Mag.  \\
\hline
2.0 & $<J-K_s$ & \\
1.5 & $<K-W1<$ & 2.5 \\
0.8 & $<W1-W2<$ & 1.8  \\
1.4 & $<W2-W3<$ & 2.4 \\
3.5 & $<W3-W4<$ & 4.5 \\
\hline
\end{tabular}
}
\end{center}
\end{minipage}
\end{table*}

\begin{table*}
\begin{minipage}{170mm}
\begin{center}
\caption{Infrared single colour cuts to select sub-mm galaxies at high redshift
}
\label{tab:table_4}
\scriptsize{
\begin{tabular}{cc}
\hline
Color & Mag.  \\
\hline
$J-K_s>$ & 2.0 \\
$K-W1>$ & 1.4 \\
$W1-W2>$ & 0.8  \\
$W2-W3<$ & 2.4 \\
$W3-W4>$ & 3.5 \\
\hline
\end{tabular}
}
\end{center}
\end{minipage}
\end{table*}

\begin{figure*}
\includegraphics[width=8cm]{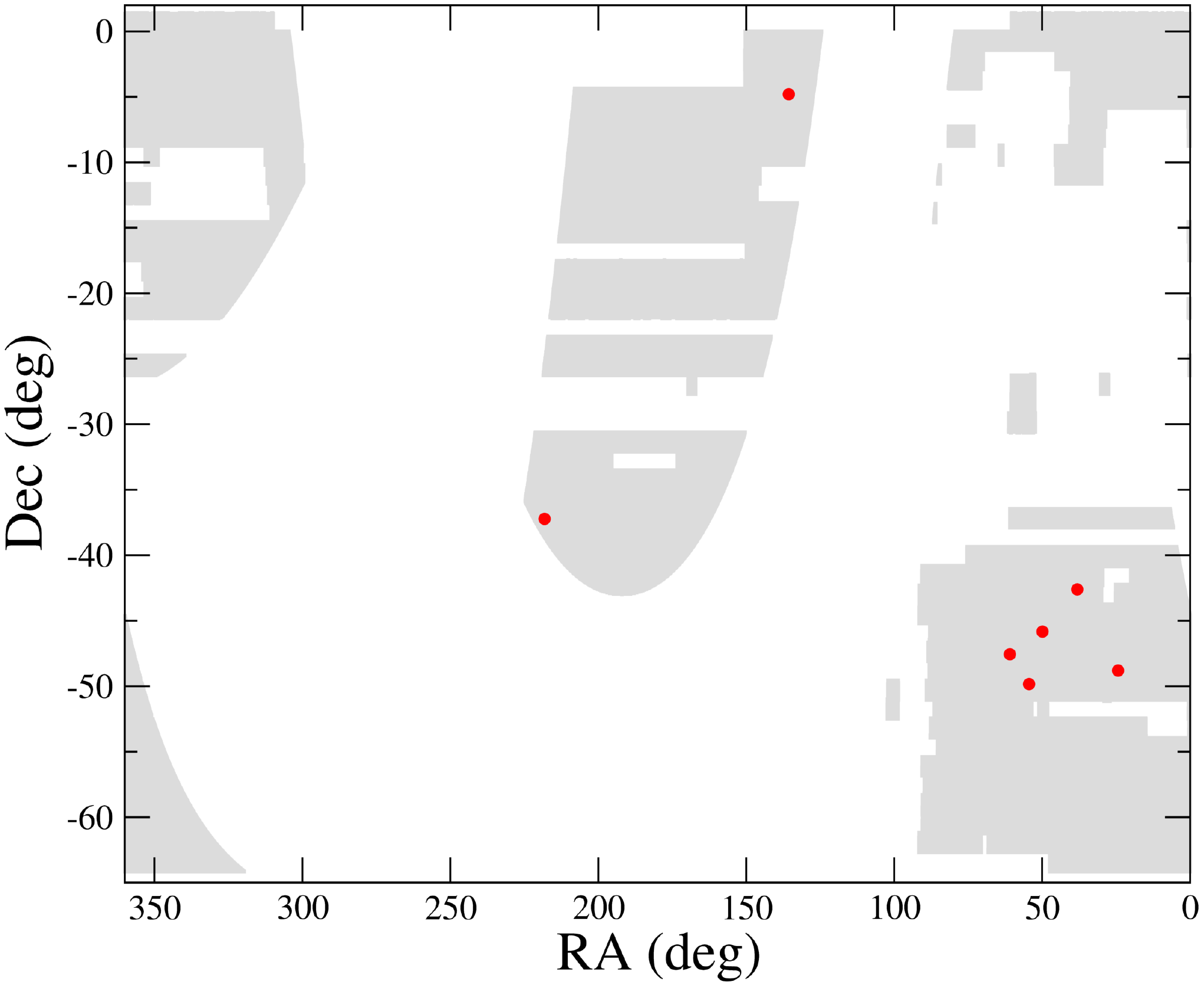}
\caption{Positions of our 7 SMG candidates (red circles) and the sky area overlapped (gray) by the  VHS/WISE survey.
}
\label{fig:f3}
\end{figure*}

\begin{figure*}
\includegraphics[width=14cm]{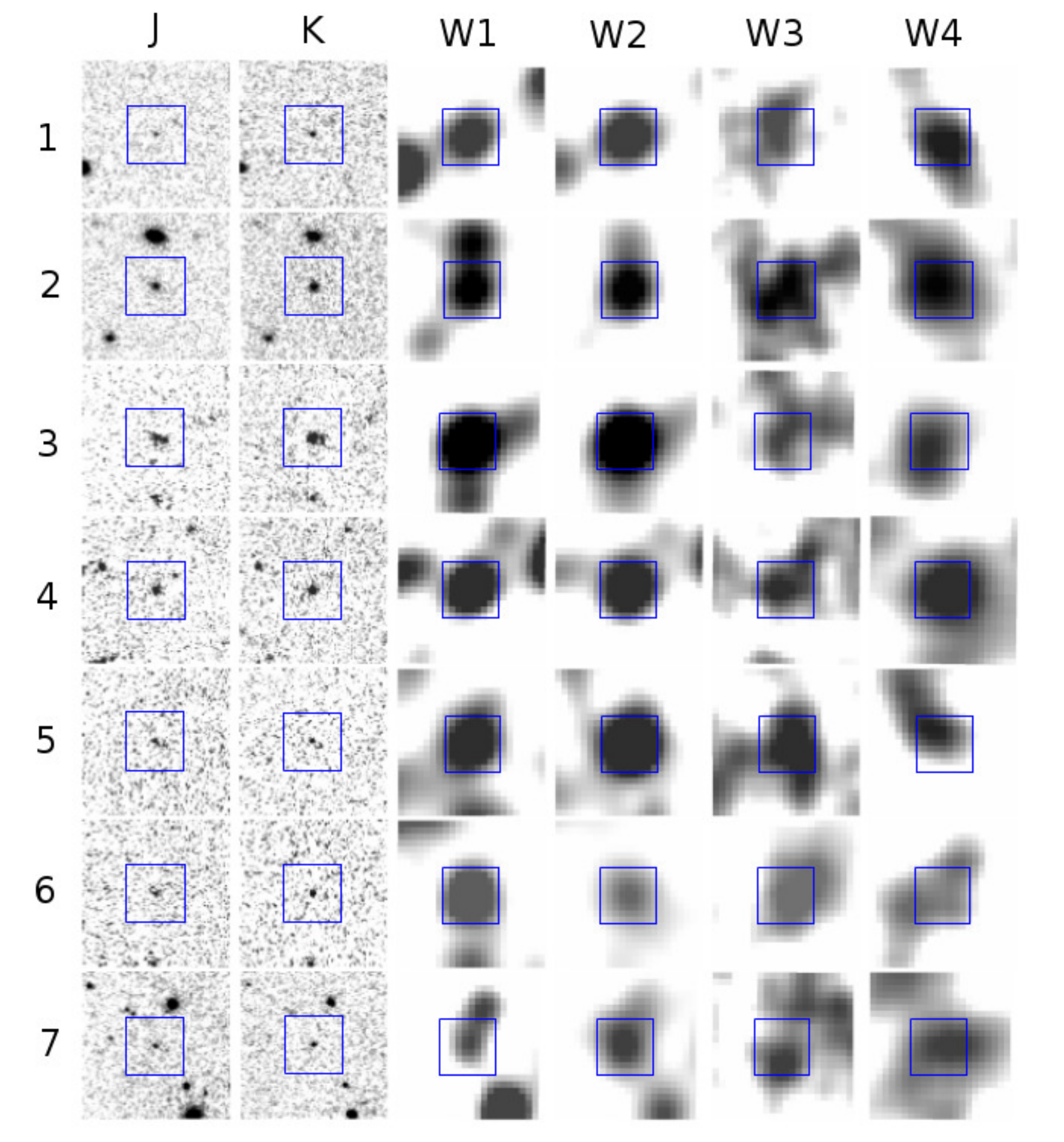}
\caption{ Images showing $0.5 \times 0.5$ arcmin$^2$ centred around candidates  in J, K$_s$, W1, W2, W3 and W4 bands. The marked boxes are $12 \times 12$ arcsec$^2$.
}
\label{fig:f4}
\end{figure*}

\begin{figure}
\begin{center}
\includegraphics[width=6.5cm]{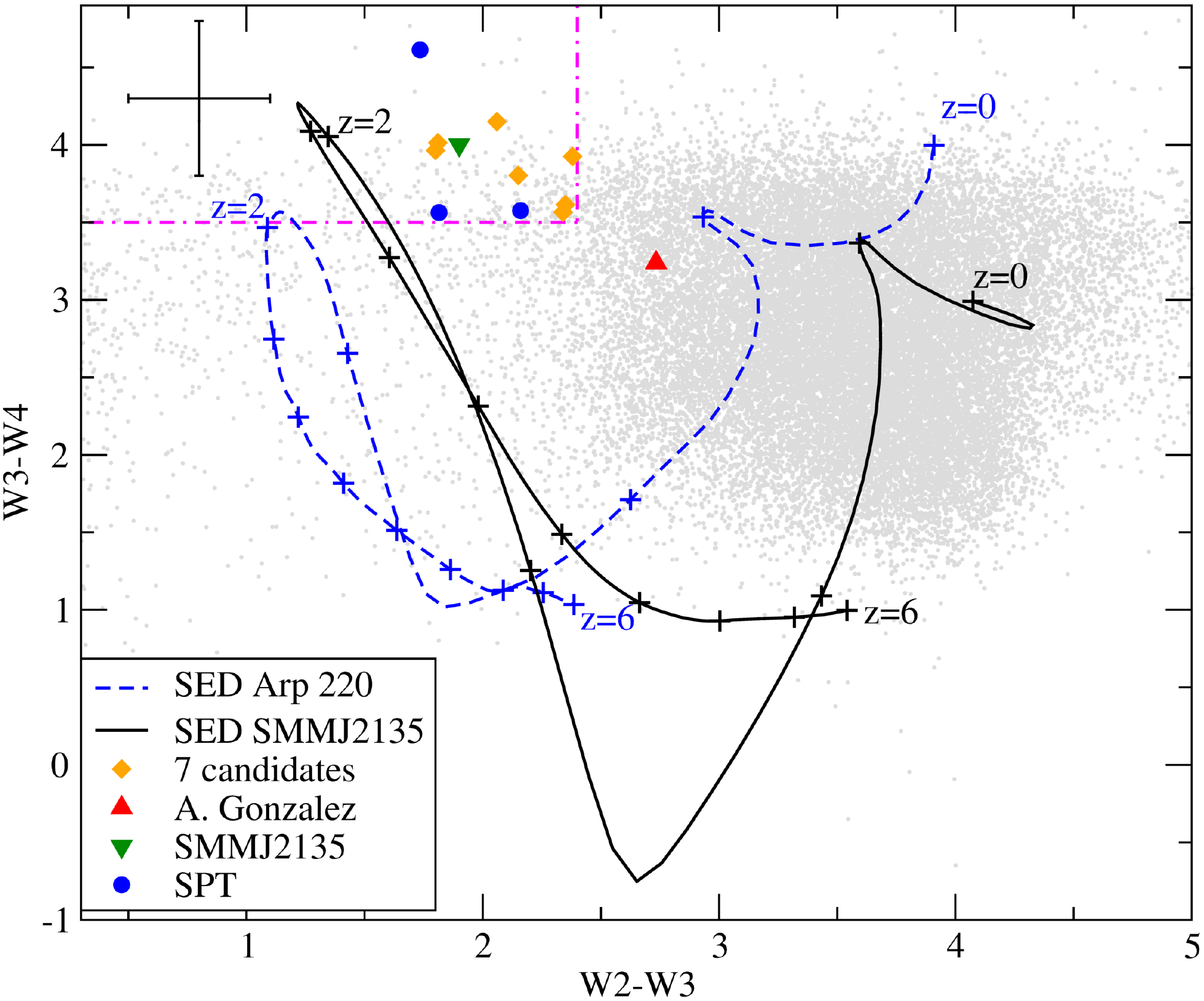}
\vspace{0.5cm}

\includegraphics[width=6.5cm]{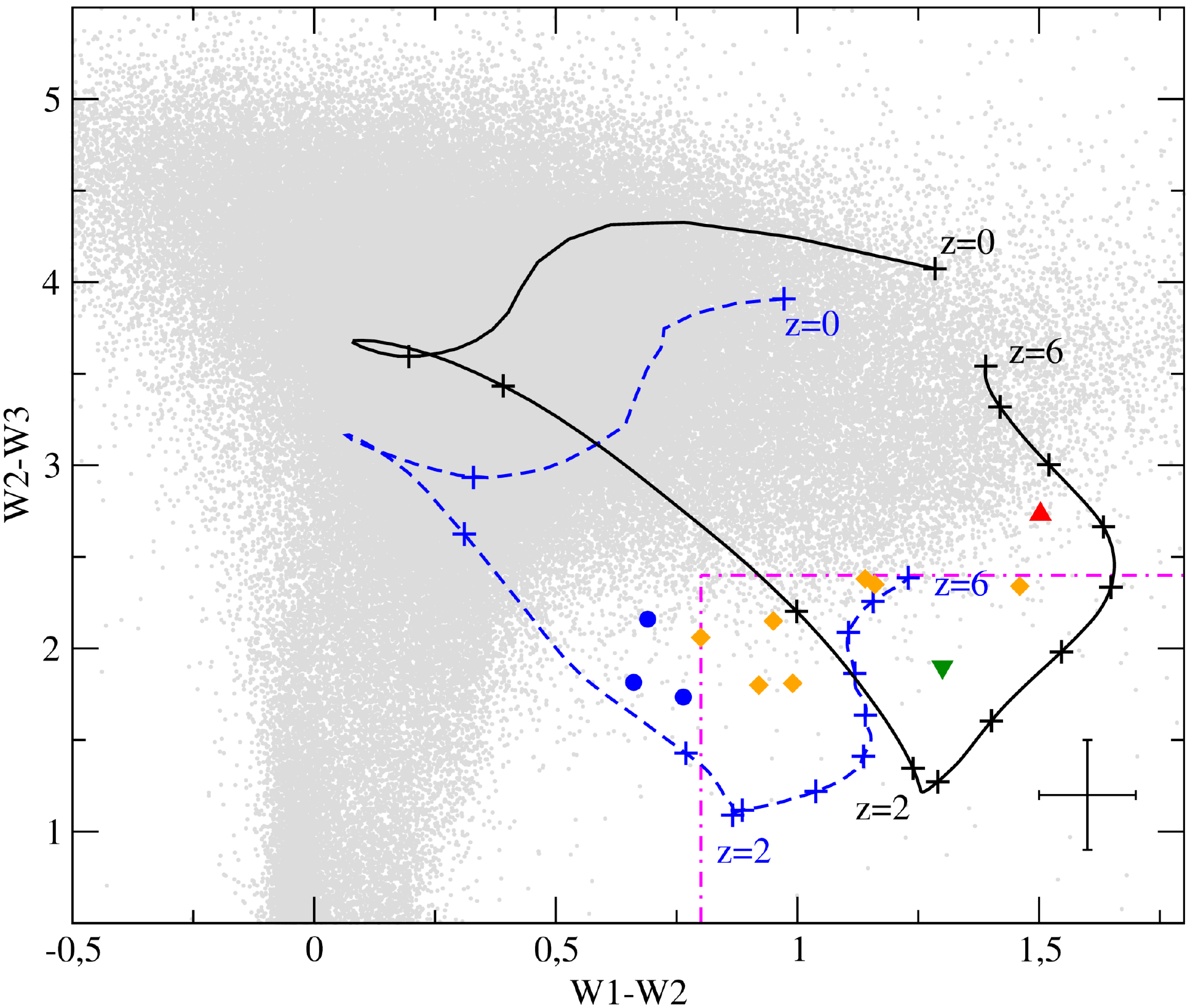}
\vspace{0.5cm}
\includegraphics[width=6.5cm]{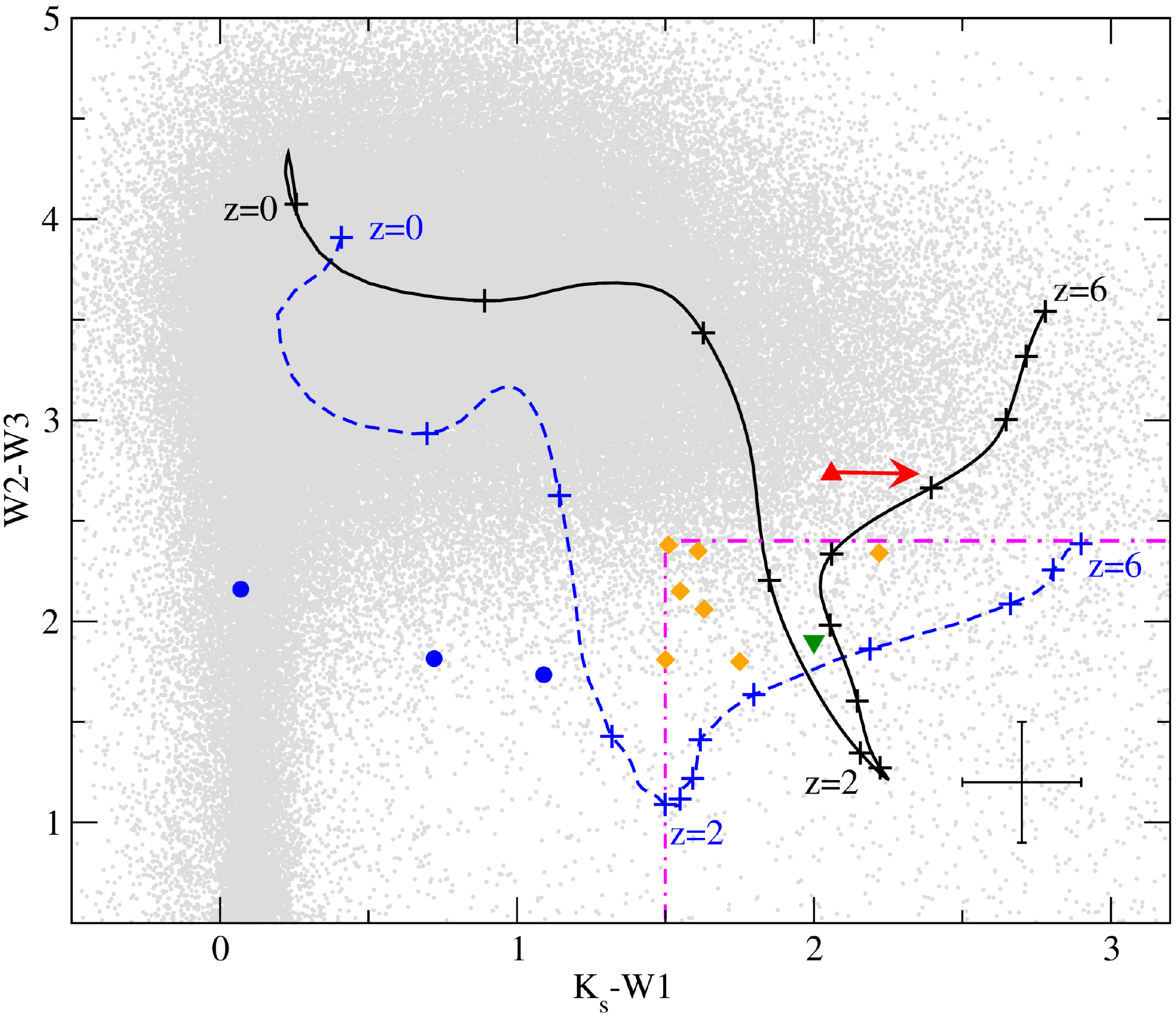}
\end{center}
\caption{Colour-colour diagrams for our 7 candidates, SMM J2135, three SPT (South Pole Telescope) sources with redshift 2.23, 2.51 and 2.78, and
 a z=2.79 SMG lensed by a galaxy cluster given in Gonzalez et al. (2011). The grey points are field sources. The colour cuts in Table 3 are given
 by point-dashed lines and typical error bars are shown in each diagram. We have used continuous lines to show the position of the SMM J2135 GRASIL SED
 and Arp220 SED as a function of redshift (indicated by a number, crosses are drawn in the curves for each 0.5 increment of redshift).}
\label{fig:f5}
\end{figure}

\subsection[]{Reference galaxy}

As a reference for our search we adopted the near and mid-IR colours of SMM J2135. This extensively studied galaxy is lensed by a cluster of galaxies with no evidence for any intervening galaxy in the line of sight which could be affecting  its near/mid-IR colours. SMM J2135 has  an intrinsic rest-frame 8-1000 $\mu$m luminosity, L$_{bol}$, of 2.3 x 10$^{12}$ L$_{\odot}$ and a star formation rate estimated at  $\sim$ 400 M$_{\odot}$/yr (Swinbank et al. 2010, Ivison et al. 2010). In the far-IR the SED can be described by dust components with temperatures of 30 and 60 K.
A comparison with local ultra-luminous infrared galaxies like Arp 220 (Ivison et al. 2010) reveals that SMM J2135 is slightly less luminous in the rest-frame optical to mid-IR (possibly due to stronger dust extinction) while  rather similar in the mm/sub-mm range.

Fig. 1 shows VHS images in J, H, K$_s$ and  WISE images in the  W1, W2, W3 and W4 bands of SMM J2135. No detection of this source was reported  in the VHS and AllWISE catalogues, however after close inspection of the relevant images, SMM J2135 can be detected in the  K$_s$-band  and in the  W3 and W4 bands. Subsequently, we measured its magnitudes  using aperture photometry  and obtained $K_s=18\pm 0.3$, $W3=12.6\pm 0.3$ and $W4=8.8 \pm 0.3$ (Vega system). An upper limit on the  J-band magnitude was also  derived from the corresponding VHS image.  The photometry was calibrated using nearby sources in the catalogue free of contaminants. In the H, W1 and W2 bands we achieve a marginal detection of SMM J2135  using the Laplacian of Gaussian filter, often called 'mexican hat', which helped to improve the detectability of sources in these filters. 
However, no reliable magnitudes could be determined from these images and  the W1 and W2 magnitudes given in Table 1 were finally estimated converting the fluxes measured  by Ivison et al in the [3.6] and [4.5] Spitzer bands,  respectively.
The  final adopted  fluxes and magnitudes of SMM J2135 are listed in Table 1. 

The fluxes of SMM J2135 from the optical to centimetre wavelengths are plotted in Fig. 2 (black dots) in comparison 
with its SED as modeled by GRASIL from Lapi et al. (2011) and  the SED of Arp 220 redshifted to z=2.32 with appropriate 
normalisation. The blue squares indicate our measurements in the K$_s$, W3 and W4 bands.

\subsection[]{Search method}

In order to find bright analogues of SMM J2135 we have correlated the AllWISE Source 
Catalogue with about 8000 $deg^2$ available from  the VHS catalogue.  The search is 
restricted to galactic latitudes $|b|\geq$20$^{o}$, reducing the surveyed area to 
6230 sq. deg., we retain galaxies for which we have detections in all the WISE bands  and identify  counterparts in the VHS catalogue within 1 arcsec of the AllWISE targets. 
 To avoid bad candidates due to Galactic extinction, the VHS and AllWISE magnitudes are 
 corrected for Galactic extinction with the use of the Schlegel et al. (1998) maps and 
 the $A_{\lambda} /E(B-V)$ coefficients taken from Cardelli et al. (1989). Then, we 
 selected those galaxies with colours consistent with the colour ranges   listed in Table 2,
  i.e. with colours within $\pm$ 0.5 mag of the SMM J2135 colours. 
 We also restricted the selection to objects with  S/N$\geq$3 in all the  WISE bands. A total of    7 galaxies verified all these conditions. 

Very recently, colour selection procedures have been refined successfully to find SMGs (see Chen et al. 2016 and references therein). These authors  
used the so-called Optical-IR Triple Colour (OIRTC) selection 
procedure. We can also adopt a similar approach, imposing just one limit for each single colour in order to select SMGs, i.e  take away one of the  
limits in each of the colours listed in   Table 2.  
When we adopt the colour limits of Table 3 we find the same candidates. Therefore we adopt for simplicity the colour criteria of Table 3 which we could call the NIR/MIR Quintuple Colour (NMIRQC) procedure.  

In Table 4 we list  photometry for the 7 resulting SMG candidates. Their sky
positions are shown in figure 3. Images of these candidates extracted from the surveys VHS and  WISE are plotted in Fig. 4.

All our candidate sub-mm galaxies have well measured positions in the NIR images (accuracy better than 0.3$^{\prime\prime}$) and are the only likely counterpart of the WISE sources. Only candidate 3 appears to be double in the J and K$_s$ images. We note that this NIR/MIR selection could be advantageous with respect to other techniques in the sub-mm range which usually require follow-up interferometry to determine subarcsec position of the candidate galaxies.

\section[]{Results and Discussion}

The 7 candidates that fulfill the colour criteria of the lensed sub-mm galaxy SMM J2135 present $J$-band magnitudes in the range 18.8-20.4 and $K_s$-band magnitudes in the range 16.8-18.2. The candidates are well identified in all bands. In spite of the different spatial resolution of the VHS and WISE, there is good agreement in the positions of the identified sources. 

In Figure 5 we plot colour-colour diagrams for the 7 candidates,  SMM J2135,  three SPT (South Pole Telescope) sources we have detected in VHS/WISE and one additional SMG lensed by a galaxy cluster reported by Gonzalez et al. 2011 (in this last case we have transformed Spitzer magnitudes to WISE magnitudes and there is a lower limit to the K$_s$  magnitude). The grey points are field sources. The colour cuts in Table 3 are given by point-dashed lines and typical error bars are shown in each diagram. We have also represented with continuous curves  the  SMM J2135 GRASIL SED and the Arp220 SED as a function of redshift (redshift is indicated by a number and crosses are drawn in the curves for each 0.5 increment of redshift).
The three SPT sources have redshift 2.23, 2.51 and 2.78, they are lensed by
foreground galaxies. While their WISE colours fall into the colour ranges we
have adopted for  SMGs, in the VHS bands there is contamination by the
intervening lensing galaxy and the K$_s$-W1 colours lie out of the adopted
range. The colours  of SMM J2135 (z=2.32) and the SMG identified by  Gonzalez et
al. 2011 (z=2.79) do probably represent the  true colours of the original lensed galaxy  because they are lensed by galaxy clusters and therefore   photometric 
contamination in the near-IR bands is not expected. From these diagrams we can see that the redshift of our selected sources are likely between z$\sim$ 1.6 and z$\sim$ 3.2. Also, it is shown that the contamination of   our sample by field sources is not large given the distant location of the adopted colour cuts with respect the bulk of the field sources. The SMM J2135 GRASIL SED does not match  well  the W3 band measurement, as we can see in figure 2, and colour differences between the SED and SMM J2135 come from this. Nevertheless the extreme values in colours W3-W4 and W2-W3 
are obtained for redshift z$\sim$ 2.5, rather far from the colour-colour regions where most field sources are located.

 In order to discuss the potential contamination of galaxies other than SMGs we have used the SWIRE template library (Polletta et al. 2007 ) which contains 25 templates including 3 ellipticals, 7 spirals, 6 starbursts (SB), 7 AGNs (3 type 1 AGNs, 4 type 2 AGNs), and 2 composite (starburst+AGN). The 3 ellipticals are 2, 5 and 13 Gyr old 
and the 7 spirals range from early to late types (S0-Sdm). 
Templates of moderately luminous AGN, representing Seyfert 1.8 and Seyfert 2
galaxies, were obtained by combining models, and spectra  of a random sample of 28 Seyfert galaxies. 
The other six AGN templates include three templates representing 
optically-selected QSOs with different values of infrared/optical flux ratios 
and two type 2 QSOs. The composite (AGN+SB) templates are empirical templates,
these objects contain a powerful starburst component, mainly
responsible for their large infrared luminosities and an AGN component that
contributes to the mid-IR luminosities.

In Figure 6 we can see that only spiral  galaxies are located in the selection region in the  $W3-W4$ vs. $W2-W3$ colour-colour diagram at redshift $z \sim 2$ (top panel) but in the  $W3-W2$ vs. $K_s-W1$ diagram 
(central panel) these galaxies only go into the selection  region for $z\sim 4$. So they do not fulfill simultaneously both colour conditions at the same redshift and therefore are not a likely source of 
contamination. The bottom panel of figure 6 shows the $W3-W4$ vs. $W2-W3$ colour-colour diagram for the 7 AGNs  and the 2 composite SB+AGN galaxies, none of them are in the selection region. The SED of SB galaxies is 
not too different from that of  Arp220 and SMM J2135 and indeed these galaxies may enter  the selection region for $1.7 \lesssim z \lesssim 3$.

In Tables 4 and 5 we complement the photometry of several SMG candidates with VHS measurements available in the H band and in the optical bands using CFHTLS. 
We have used redshifted  GRASIL SEDs of SMM J2135 to determine the photometric
redshift of our candidates, from  J,H,K$_s$,W1,W2,W3,W4-band measurements (H-band when available) using a standard $\chi^2$ minimization procedure. The results  are given in Table 4. All derived redshifts lie in the range z=1.6-3.2 as expected. We find 2 galaxies with K$_s$ $\leq$ 16.8, 3 galaxies in the range K$_s$=17-17.2
and 2 galaxies with K$_s$=17.5-18.2  Only 1 galaxy has photometric redshifts higher than  z=3 in our sample. 
In general our selected candidates are about 1-2 magnitudes brighter than
typical lensed SMGs selected from Herschel surveys (e.g. Ma et al. 2015a; Calanog et al. 2014) or the SPT survey (Ma et al. 2015b). However, one source, HATLAS J142935.3-002836 alias G15v2.19, at z=1.027 has similar NIR properties as our brightest candidates (Messias et al. 2014; Calanog et al. 2014; Ma et al. 2014).

In Fig. 7 we plot the SEDs of the candidate SMGs in comparison with the GRASIL SED of SMM J2135 shifted to the photometric redshift determined for  
each candidate. We note that in most cases the model fits very well all the near-IR and mid-IR bands. In the case of candidate 6 (VHSJ0902-0448), 
however, the fluxes deviate from the model at the shorter wavelength bands, probably because of the presence of an intervening galaxy in the line of 
sight which however is not detected in the available J and K$_s$ images. From these fits we can estimate peak sub-mm fluxes above $\sim$500 micron and a flux density at 1.4 mm above $\sim$ 15 mJy 
for all of our targets. Very few sources are expected to be intrinsically bright enough to exceed these fluxes (Karim et al. 2013) and therefore we may expect that our sample is conformed by gravitationally lensed galaxies. 

 Weiss et al. (2013)  measured the redshift distribution of 26 strongly lensed sources selected from the 2500 square-degrees of the South Pole Telescope (SPT) survey with fluxes at 
 1.4mm above 20 mJy. They found a redshift distribution of the SPT sample with a mean of z=2.0,  only 1 out of their 26 targets had redshift in the range z=2.5-3.2. Our targets have redshifts 
 consistent with the SPT distribution, but only for 5 targets we estimate 1.4 mm fluxes above 20 mJy from the fits plotted in Fig. 7. The surface density we infer for SMG sources selected in the 
 near/mid IR  to match 1.4 mm fluxes above 20 mJy would be 5/6230 deg$^{-1}$ i.e. approximately 10 times less  surface density that that found by Weiss et al. (2013) for potentially equivalent 
 objects. We argue that this difference  can be due to our selection criteria mostly identifying galaxies which are  lensed by clusters of galaxies. Galaxies amplified by an intervening galaxy 
 in the line of sight are likely to have photometric magnitudes in the J and K$_s$ bands affected by the lens and may not  verify our selection criteria. Only a few cases in SPT are cluster lensing galaxies. Spilker et al. (2016) find that 4 of 
 47 SPT sources are lensed by clusters.
 The NMIRQC procedure, proposed here, could  be a very effective tool to identify strongly  lensed SMGs  by clusters of galaxies. SMGs detected by Herschel have 500 $\mu$m fluxes in the range 100-350 mJy (Bussmann et 
 al. 2013). For our candidates, we estimated  500 $\mu$m fluxes from the fitted SEDs, the results range from 160 to 980 mJy and are listed in Table 4.  Follow-up sub-mm/mm observations of the candidates will demonstrate if indeed we are dealing with strongly lensed SMGs.

From the same SEDs we obtained  60 and 100 $\mu$m fluxes  which used with    the FIR/radio relation  by Condon et al. (1992) allowed us to  predict radio emission at 1.4 GHz in the range 0.09-0.36 mJy. To our knowledge no detection of radio emission  has been reported in the  literature for our targets.

\subsection[]{Clusters of galaxies and sub-mm galaxy candidates}

Using the red sequence method (Gladders et al. 2000) and the VHS survey we have searched for clusters of galaxies near our  SMG candidates and identified several galaxy clusters in the range $0.2 < z < 0.9$. 

We searched for local over densities of galaxies using the $J-K_s$ versus $J$ colour-magnitude diagram. For these redshifts, the slope of the red-sequence in the 
diagram is nearly $-0.03$ (Cohn et al., 2007). We took $J-K_s$ colour slices of amplitude $0.2$ magnitudes and searched for local overdensities in circular areas around the brightest cluster galaxy (BCG) with radius 
$\sim 1$ Mpc.
We selected a cluster candidate if the density of galaxies of the cluster were larger than 3 times the mean density and had  more than 10 galaxies in the red-sequence. We adopted as cluster radius the largest radius where the over density was detected.

In Figure 8 we can see the colour-magnitude diagrams for the six clusters of galaxies we have detected near SMG candidates 1, 2, 4, 5, 6, and 7, respectively. The $J-K_s$ colour of the BCG is an indicator of the 
redshift of the cluster (Tonini et al 2012). Only for candidate 3 we do not find a nearby cluster. Indeed this candidate is the only one that appears to be double in the VHS images of Figure 4, suggesting the presence of an intervening galaxy in the line of sight.
We estimated the redshift of the clusters from the colours predicted by the red sequence fit (see Fig. 8) for each BCG. 
In Table 6 we present the main properties of all these clusters and group of galaxies, including reacheness and radius estimates. 
The VHS $J$-band images $10\times10$ arcmin$^2$ are shown in Fig. 9, blue circles denote clusters and are  plotted with the cluster radii given in Table 6. The blue boxes mark the position of SMG candidates. Red circles denote the 
position of identified candidate galaxy members. Remarkablly  candidates 1, 2 and 4 are located within  clusters and candidate 5 is just at 1.8 arcmin from the centre of a cluster. The identification of clusters of galaxies within 3.5 arcmin of six out of 7  
candidates further  supports the suggestion that these objects are lensed  SMGs.

\begin{figure*}
\begin{center}
\includegraphics[width=6.5cm]{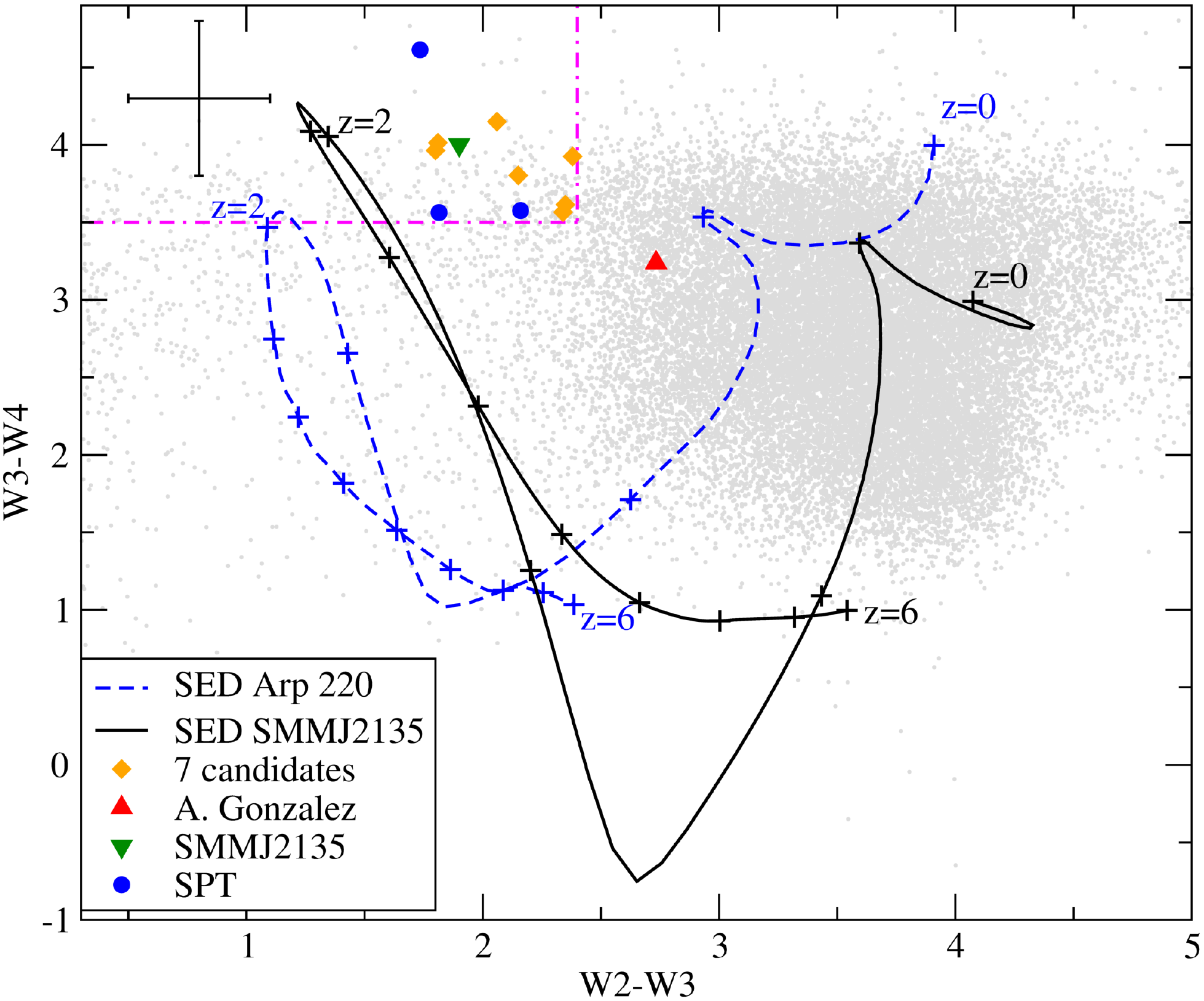}
\vspace{0.5cm}

\includegraphics[width=6.5cm]{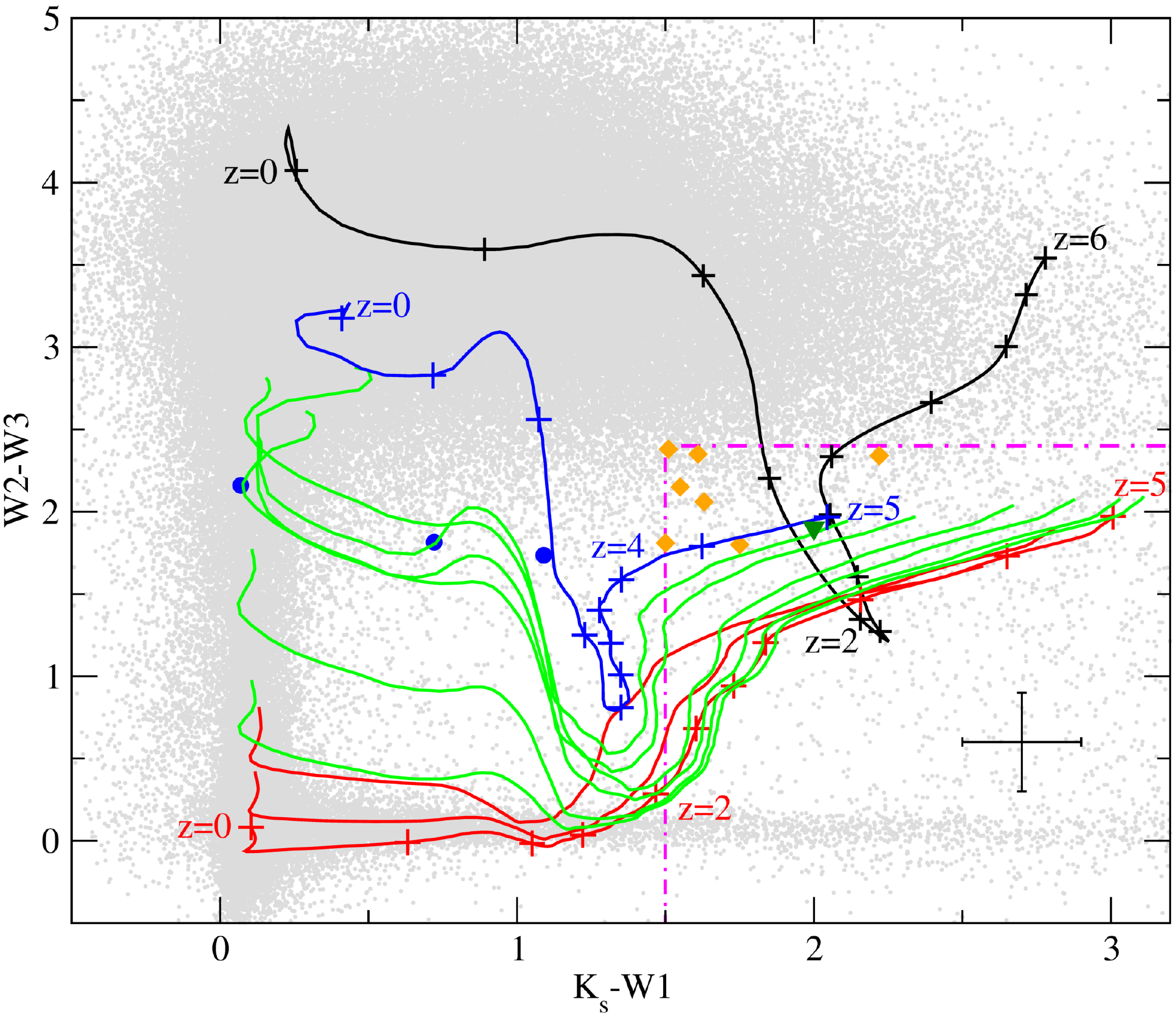}
\vspace{0.5cm}

\includegraphics[width=6.5cm]{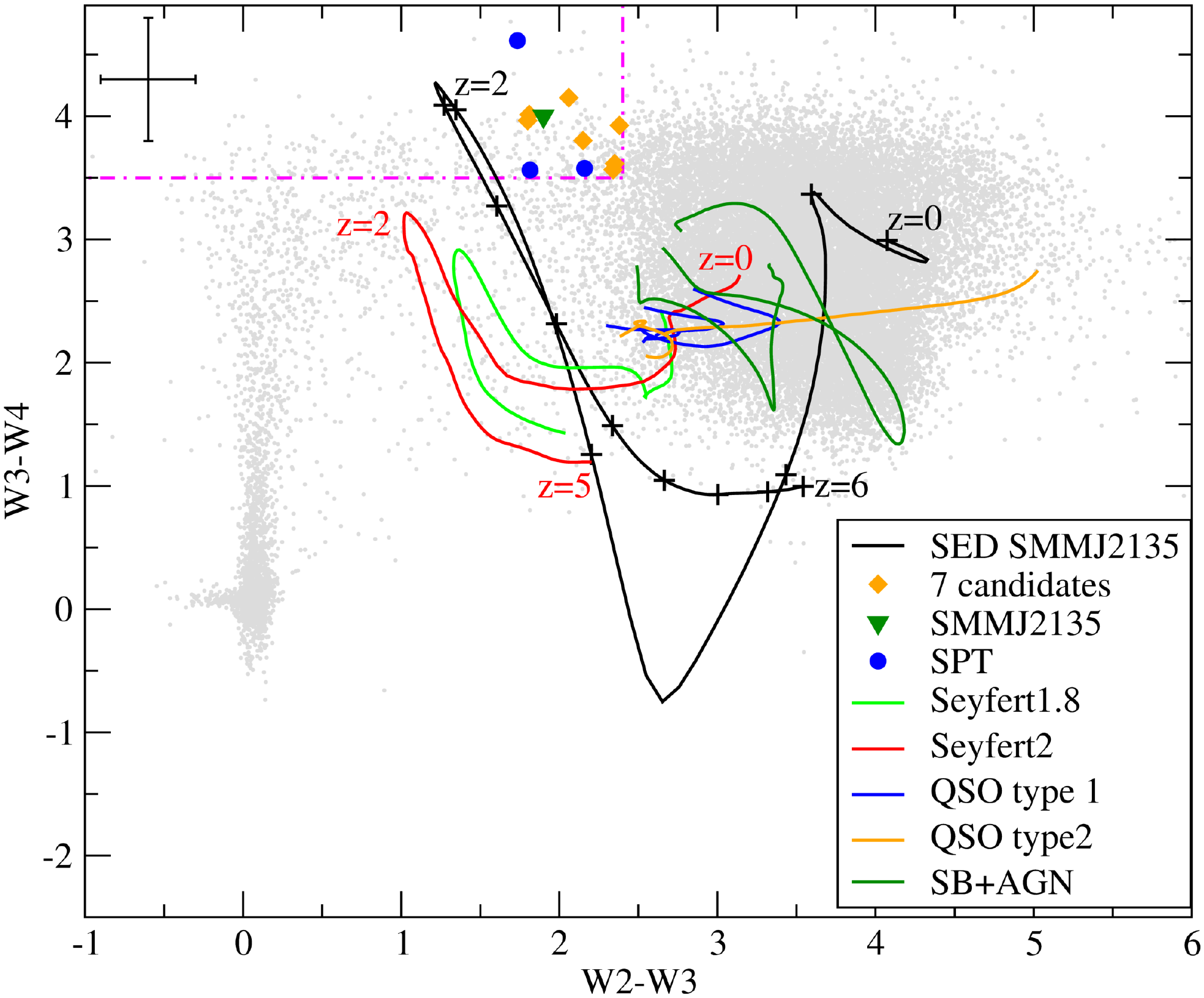}
\end{center}
\caption{Colour-colour diagrams for our 7 candidates, SMM J2135 and three SPT (South Pole Telescope) sources with redshift 2.23, 2.51 and 2.78.
The grey points are field sources. The color cuts in Table 3 are given by point-dashed lines and tipical error bars are shown in each diagram. 
Top and central panels: the continuous curves show the position of SMM J2135 GRASIL SED, 3 elliptical galaxies which are 2, 5 and 13 Gyr old 
and 7 spirals range from early to late types (S0-Sdm). Bottom panel: the continuous curves show the position of SMM J2135 GRASIL SED, 
2 moderately luminous AGN, representing Seyfert 1.8 and Seyfert 2 galaxies, 3 type 1 QSOs, 2 type 1 QSOs and 2 composite starburst(SB)+AGN galaxies  
(in some curves redshift is indicated by a number in the same colour and crosses are drawn in these curves for each 0.5 increment of redshift).}
\label{fig:f6}
\end{figure*}

\begin{figure*}
\includegraphics[width=16cm]{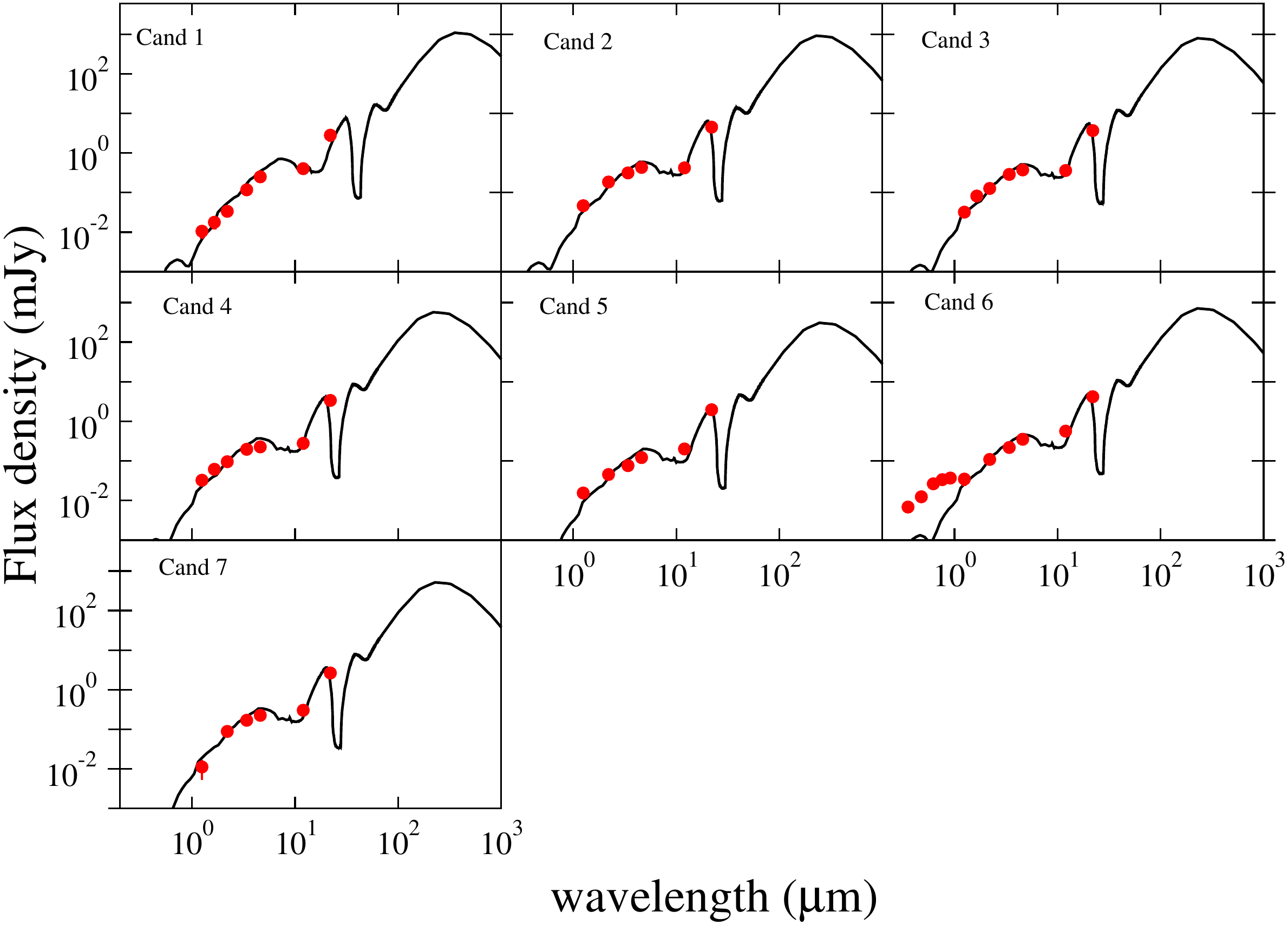}
\caption{Fluxes of the SMG candidates (red points) in comparison with the SMM J2135 GRASIL SED shifted to the photometric redshift of each candidate (black lines).
Note that only candidate ID1 has values greater than 500mJy (980mJy) at 500 microns. The rest have values between 160 and 430 mJy}
\label{fig:f7}
\end{figure*}

\begin{figure*}
\includegraphics[width=16cm]{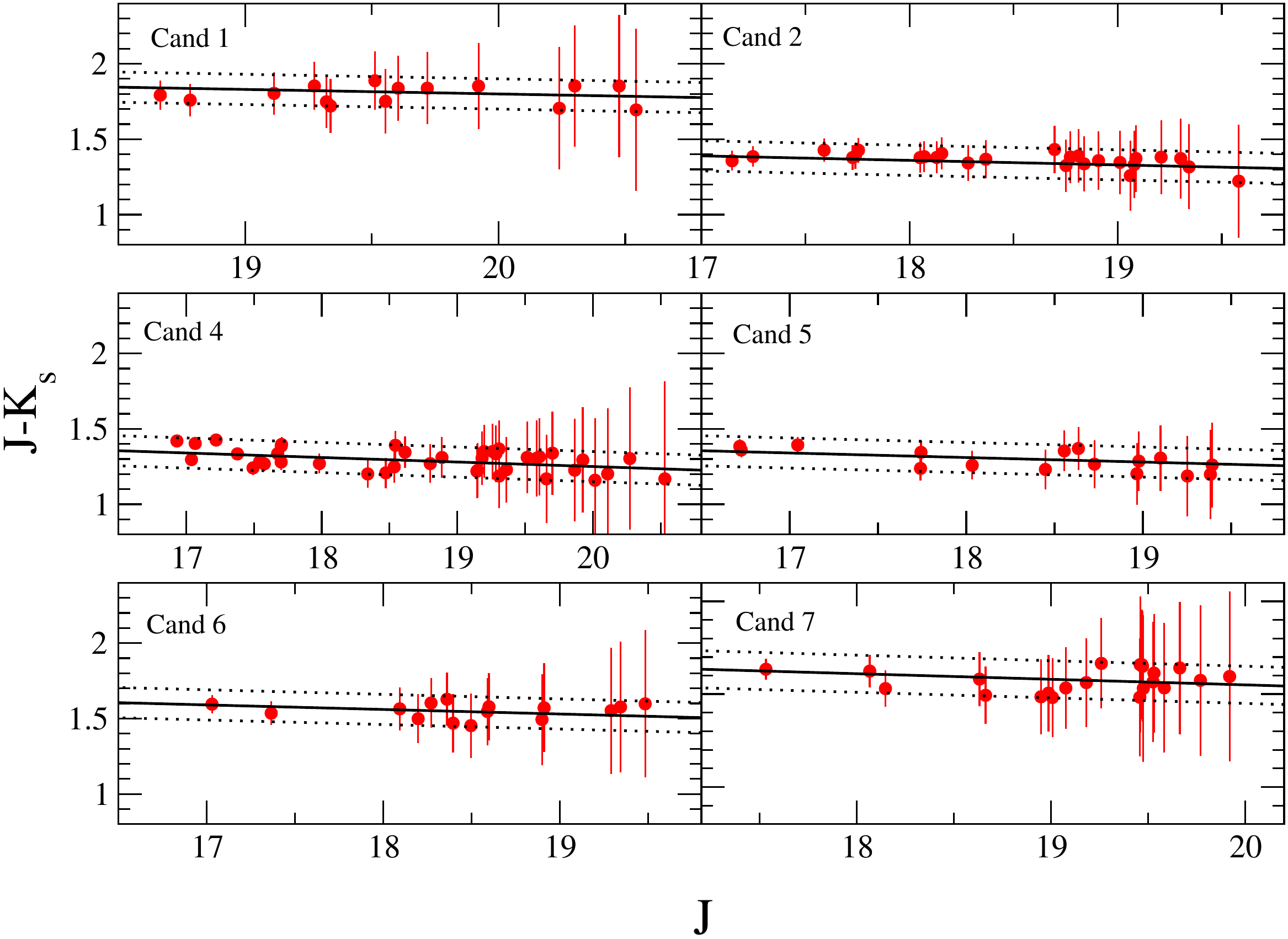}
\caption{
$J-K_s$ vs. $J$ colour-magnitude diagrams for 6 galaxy clusters near our  SMG candidates (number labelled  on each diagram).
Diagrams show the objects within 1 Mpc from the center of each galaxy cluster. The slope of the red sequence is $-0.03$ and the dotted
lines are located at a separation of 0.1 magnitudes from the red sequence line.
}
\label{fig:f8}
\end{figure*}

\begin{figure*}
\includegraphics[width=16cm]{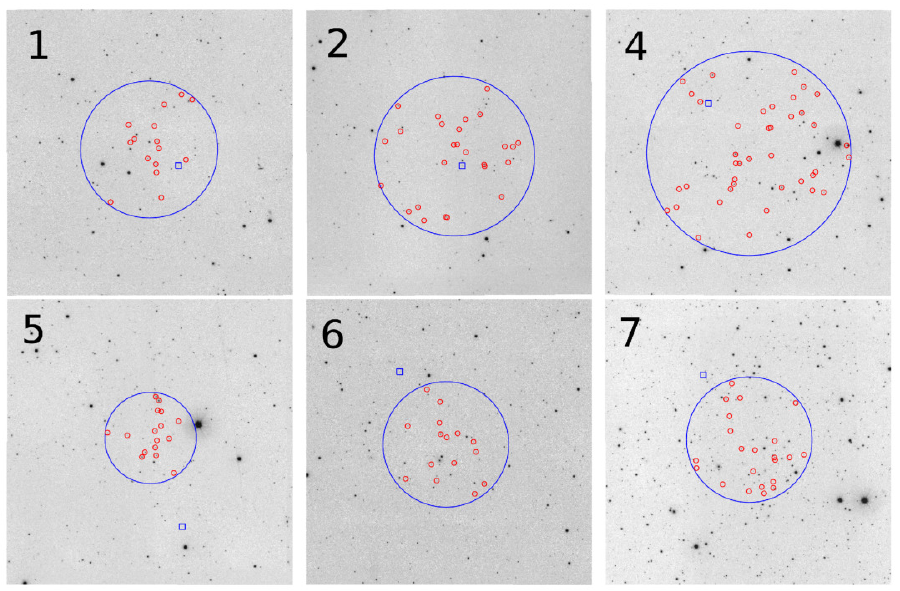}
\caption{$J$-band images of 6 galaxy clusters identified near our SMG candidates (labelled with the corresponding number on each image). Blue circles mark the core of the cluster. Red circles mark galaxies in the red 
sequence of the cluster. Blue boxes mark the  positions of SMG candidates.}
\label{fig:f9}
\end{figure*}

\begin{table*}
\begin{minipage}{170mm}
\begin{center}
\caption{ VISTA (http://horus.roe.ac.uk/vsa/index.html) and WISE magnitudes.}
\scriptsize{
\begin{tabular}{ccccccccccc}
\hline
 & &   &  &  &  &  &  &  & & \\
ID & RA & Dec & $J$  & $H$   & $K_s$   & $W1$  & $W2$  & $W3$  & $W4$ & $z_{phot}$  \\

 & &   &  &  &  &  &  &  & & \\
\hline
1 VHSJ0137-4848  & 01:37:24.39 & -48:48:25.1 & 20.4 $\pm 0.4$ & 19.4  $\pm 0.4$  & 18.2 $\pm 0.3$ & 16.03 $\pm 0.05$ & 14.56 $\pm 0.05$ & 12.2 $\pm 0.3$ & 8.7 $\pm 0.4$ & 3.2 $\pm 0.2$ \\ 
2 VHSJ0232-4236 & 02:32:37.00 & -42:36:31.2 & 18.8 $\pm 0.2$ &                  & 16.5 $\pm 0.1$ & 14.96 $\pm 0.03$ & 13.97 $\pm 0.04$ & 12.2 $\pm 0.3$ & 8.2 $\pm 0.3$ & 1.7 $\pm 0.1$ \\ 
3 VHSJ0319-4549& 03:19:38.74 & -45:49:40.2 & 19.2 $\pm 0.2$ & 17.7  $\pm 0.2$  & 16.8 $\pm 0.2$ & 15.06 $\pm 0.03$ & 14.13 $\pm 0.03$ & 12.3 $\pm 0.3$ & 8.4 $\pm 0.2$ & 1.7 $\pm 0.1$ \\ 
4 VHSJ0337-4950& 03:37:46.05 & -49:50:08.7 & 19.2 $\pm 0.2$ & 18.0  $\pm 0.2$  & 17.1 $\pm 0.2$ & 15.47 $\pm 0.03$ & 14.68 $\pm 0.04$ & 12.6 $\pm 0.3$ & 8.5 $\pm 0.2$ & 1.6 $\pm 0.1$ \\ 
5 VHSJ0403-4733 & 04:03:39.52 & -47:33:37.4 & 20.0 $\pm 0.3$ &                  & 18.0 $\pm 0.3$ & 16.49 $\pm 0.05$ & 15.35 $\pm 0.06$ & 13.0 $\pm 0.3$ & 9.0 $\pm 0.3$ & 1.9 $\pm 0.1$ \\ 
6 VHSJ0902-0448& 09:02:23.98 & -04:48:17.7 & 19.1 $\pm 0.3$ &                  & 17.0 $\pm 0.2$ & 15.35 $\pm 0.04$ & 14.20 $\pm 0.05$ & 11.9 $\pm 0.3$ & 8.2 $\pm 0.3$ & 1.9 $\pm 0.1$ \\ 
7 VHSJ1432-3713& 14:32:41.33 & -37:13:40.6 & 20.3 $\pm 0.8$ &                  & 17.2 $\pm 0.2$ & 15.63 $\pm 0.05$ & 14.68 $\pm 0.06$ & 12.5 $\pm 0.3$ & 8.7 $\pm 0.3$ & 1.7 $\pm 0.1$ \\ 
\hline
\end{tabular}
}

\end{center}
\end{minipage}
\end{table*}

\begin{table*}
\caption{
Optical magnitudes.
}
\begin{center}
\begin{tabular}{|c|c|c|c|c|c|c|}
\hline
 &  &  &  &  & & \\
ID  & $u$  & $g$  &  $r$  & $i$ &  $z$  & Survey  \\
 &  &  &  &  & & \\
\hline
6 VHSJ0319-4549 & 21.81 $\pm 0.01$  & 21.161 $\pm 0.004$ & 20.341 $\pm 0.003$ & 20.083 $\pm 0.003$ & 19.984 $\pm 0.006$ & CFHTLS \\
\hline
\end{tabular}
\end{center}
\end{table*}

\begin{table*}
\begin{center}
\caption{Cluster of galaxies near some candidates}
\scriptsize{
\begin{tabular}{cccccc}
\hline
  & & & & & \\
Candidate & Redshift$^1$ & Galaxies$^2$ & Cluster radius$^3$ & Distance$^4$ & Overdensity factor$^5$ \\
  & & & & &  \\
\hline
1 VHSJ0137-4848 & 0.9 & 15 & 2.4 & 1.10 & 3.72 \\
2 VHSJ0232-4236 & 0.3 & 26 & 2.8 & 0.43 & 3.11 \\
4 VHSJ0337-4950 & 0.2 & 40 & 3.5 & 2.44 & 3.39 \\
5 VHSJ0403-4733  & 0.2 & 16 & 1.6 & 3.43 & 4.32 \\
6 VHSJ0902-0448 & 0.5 & 15 & 2.2 & 3.20 & 3.62 \\
7 VHSJ1432-3713 & 0.5 & 22 & 2.2 & 3.16 & 3.67 \\
\hline
\end{tabular}
}

$^1$ Estimate from the $J-K_s$ colour of the BCG and figure 3 in \cite{CS09,TB12}.

$^2$ Number of galaxies in the red-sequence of the cluster.

$^3$ The largest radius where the overdensity is detected in arcmin.

$^4$ Projected distance of the SMG to the geometric center of the cluster in arcmin.

$^5$ Overdensity factor from the mean denstity.

\end{center}
\end{table*}

\section[]{Conclusions}

We report results on a search for bright high redshift  submillimetre galaxies , analogues of the lensed galaxy SMM J2135-0102, z=2.3, via a cross-correlation of the WISE and VHS databases performed over a high galactic latitude southern sky area of  6230 sq deg. We adopted as reference the near/mid-IR colours of this galaxy and searched for galaxies of similar colours with reported detections in  all  J, K$_s$, W1, W2, W3 and W4 bands. We find 7 galaxies brighter than K$_s$=18.2 matching the adopted colour criteria. 

Using  redshifted SEDs of the reference galaxy we determine the most likely redshift for each galaxy and find that our sample lie in the range z=1.6-3.2. From the best individual fits we estimate sub-mm and mm fluxes for each target and conclude that 5 out of the 7 galaxies may have fluxes above 20 mJy  at 1.4 mm. The surface density that we determine for lensed SMGs ($\sim$ 1 galaxy per 890 sq. deg.) is about 10 times lower than that obtained by Weiss et al. (2013) for z=1.6-3.2 sub-mm galaxies with 1.4 mm fluxes above 20 mJy. The difference can be due to our selection criteria mostly  selecting SMGs amplified by clusters of galaxies, as those amplified by intervening galaxies may have their infrared colours affected.  

Using VHS and the J-K vs J cluster sequence method,  we have identified potential clusters/groups of galaxies near 6 of these candidates. Photometric redshifts locate most of these nearby clusters in the range z=0.2-0.9. 

 We propose that our near/mid-IR selection procedure can identify good candidates to  bright high redshift lensed SMGs. Follow-up sub-mm observations (already on-going with APEX) and mm-observations (with e.g. LMT and ALMA)  will reveal if indeed the selected objects are bright lensed SMGs and will provide accurate spectral redshifts. If confirmed, these would be excellent targets to carry out a systematic determination of the properties in the optical, near-IR, mid-IR, millimetre and submillimetre of SMGs in the redshift range 1.6-3.2. 

The near/mid IR colour selection used in this work can potentially be extended in 
redshift space, particularly to higher redshifts where the redshift distribution of SMGs displays a peak (e.g. B\`ethermin et al. 2015). Higher redshift searches increasingly demand higher sensitivity near-IR data. Future large scale deep near-IR surveys, as  planned for the Euclid mission (ESA), in combination with WISE  may enable the detection of a large population of bright lensed SMGs at various redshifts.  

\section*{Acknowledgements}

This work has been partially funded by projects "Participacion en el instrumento NISP y preparacion para la ciencia de EUCLID", ESP2015-69020-C2-1-R (MINECO). 
H.D. acknowledges financial support from the Spanish Ministry of Economy and Competitiveness (MINECO) under the 2014 Ram\'on y Cajal program MINECO
 RYC-2014-15686

\label{lastpage}

\end{document}